\documentclass{aa} 
\usepackage{graphicx}
\usepackage{txfonts}
\usepackage{latexsym}
\usepackage{natbib} 
\usepackage{amssymb,mathabx}
\bibpunct{(}{)}{;}{a}{}{,}
\usepackage{xcolor}

\definecolor{beige}{HTML}{FFDF8C}

\begin{document}
\title{Variable rotational line broadening in the Be star Achernar\thanks
  {Based on observations collected at the European Southern Observatory at La
    Silla and Paranal, Chile, Prog.\ IDs: 62.H-0319, 64.H-0548, 072.C-0513,
    073.C-0784, 074.C-0012, 073.D-0547, 076.C-0431, 077.D-0390, 077.D-0605,
    and the technical program IDs 60.A-9120 and 60.A-9036.}}
%
%
\author{Th.~Rivinius\inst{1} \and D.~Baade\inst{2} \and R.~H.~D.~Townsend\inst{3}
\and A.~C.~Carciofi\inst{4} \and S.~\v{S}tefl\inst{5}}
%
%
%
\institute{ESO --- European Organisation for Astron.\ Research in the
  Southern Hemisphere, Santiago, Chile,  Casilla 19001
\and
ESO --- European Organisation for Astron.\ Research in the
  Southern Hemisphere, Garching, Germany, Karl-Schwarzschild-Str. 2
\and 
Department of Astronomy, Univ.\ of Wisconsin-Madison, 2535 Sterling Hall, 475 N. Charter Street, Madison, WI 53706, USA
\and 
Instituto de Astronomia, Geof\'isica e Ci\^encias Atmosf\'ericas, Universidade
de S\~ao Paulo, 05508-900, S\~ao Paulo, SP, Brazil
%
%
\and ESO/ALMA --- The Atacama Large Millimeter/Submillimeter Array, Alonso
de C\'ordova 3107, Vitacura, Santiago, Chile}
\date{Received: $<$date$>$; accepted: $<$date$>$; \LaTeX ed: \today}
\abstract%
{}
{The main theoretical problem for the formation of a Keplerian disk around Be
  stars is how to supply angular momentum from the star to the disk, even more
  so since Be stars probably rotate somewhat sub-critically. For instance,
  nonradial pulsation may transport angular momentum to the stellar surface until
  (part of) this excess supports the disk formation/replenishment. The nearby
  Be star Achernar is presently building a new disk and offers an excellent
  opportunity to observe this process from relatively close-up.  }
{Spectra from various sources and epochs are scrutinized to
  identify the salient stellar parameters characterizing the disk life cycle
  as defined by H$\alpha$ emission.  Variable strength of the non-radial
  pulsation is confirmed, but does not affect the further results.}
{For the first time it is demonstrated that the photospheric line width does
  vary in a Be star, by as much as $\Delta v \sin i \lesssim
  35$\,km\,s$^{-1}$.  However, contrary to assumptions in which a photospheric
  spin-up accumulates during the diskless phase and then is released into the
  disk as it is fed, the apparent photospheric spin-up is positively
  correlated with the appearance of H$\alpha$ line emission: The photospheric
  line widths and circumstellar emission increase together, and the apparent
  stellar rotation declines to the value at quiescence after the H$\alpha$
  line emission becomes undetectable.}
{}
\maketitle
%

\section{Introduction}  

Be stars are rapidly rotating stars that form gaseous, circumstellar viscous
decretion disks \citep[see][for a review]{review}. In order to form a viscous
decretion disk and keep it in existence, angular momentum must be supplied to
the inner part of the disk, which is then transported outwards by viscosity
\citep[see, e.g.,][]{2012ApJ...756..156H}.  It is generally considered that Be
stars rotate close to, but not at critical value, so that there has to be a
disk formation process acting on top of the rotation. Since in many Be stars
the disk is transient, either the formation or the dissipation process of the
disk, or both, must be able to vary with time in efficiency for a given star.
Among the proposed mechanisms for this is the pulsational build-up of excess
angular momentum in the upper photosphere that eventually ``offloads'' into a
circumstellar disk
\citep[][]{1986A&A...163...97A,2013ApJ...772...21R,Coralie_2013}. Such a
mechanism should leave an imprint on the photosphere by increasing the
rotational line widths, usually expressed as $v \sin i$, {\em before} the disk
is formed/supplied with angular momentum, and then should decrease as the disk
forms. To our knowledge, no observational results from searching for such a
phenomenon have been published.

The southern Be star \object{Achernar} ($\alpha$\,Eri, HD\,10144) is the
apparently brightest Be star.  The star is mostly classified as B3\,V or
B4\,IV, but sometimes as well as B6\,V.  \citet{2006A&A...446..643V} attempted
to reconcile these discrepant spectral types and other published parameter
values by taking the rapid rotation of the star into account, and derived a
``surface mean effective temperature'' of 15\,000\,K. Achernar's rapid
rotation is typical of Be stars, but it is the only Be star for which the
rotational flattening has actually been observed
\citep{2003A&A...407L..47D}. \citet{2012A&A...545A.130D} revisited the star,
and based on more interferometric data confirmed the initial findings of
Achernar rotating at about 95\% of the critical velocity. However, the
determinations of the actual projected rotational speed differ quite widely:
\citet{2006A&A...446..643V} obtained $v \sin i = 223\pm15\,{\rm km\,s^{-1}}$,
while \citet{2012A&A...545A.130D} give $292\pm10\,{\rm km\,s^{-1}}$.

\begin{figure*}[t]
\begin{center}
\includegraphics[angle=0,width=18cm,clip]{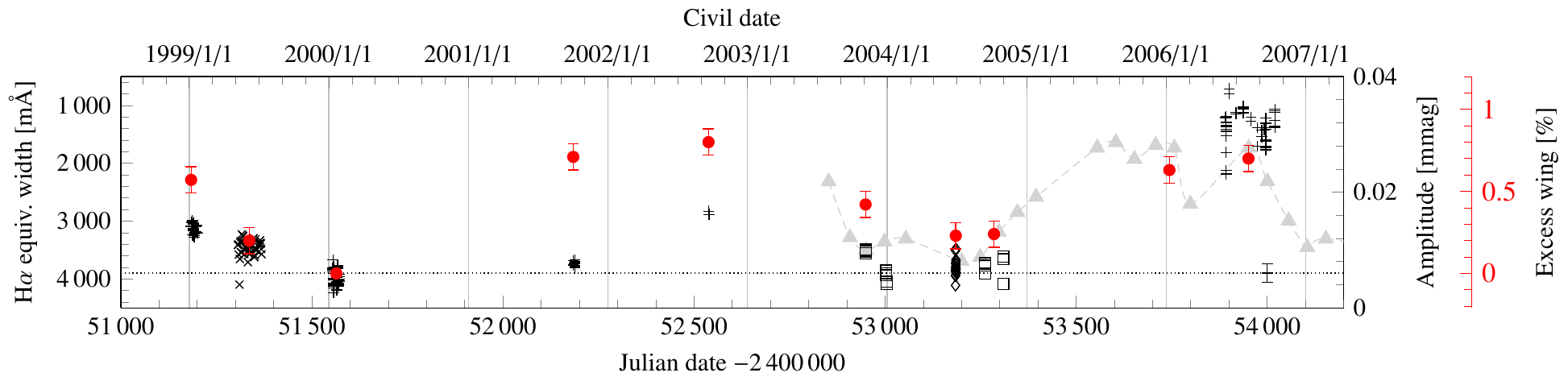}%
\end{center}
\caption[xx]{\label{fig:EW} Achernar H$\alpha$ equivalent width 1999 to 2007.
  + for FEROS, $\times$ for HEROS, \raisebox{.2ex}{\scalebox{0.7}{$\Box$}} for
  HARPS, and {\large$\diamond$} for UVES. The mean value in diskless state,
  3900\,m\AA, is indicated as dotted line, the mean uncertainty of
  $\pm160$\,m\AA, estimated as RMS in diskless state, is shown exemplarily in
  the lower right corner. Light grey triangles
  ({\Large\raisebox{-.3ex}{$\blacktriangleup$}}) are photometric amplitudes of
  frequency $F_1$ taken from Fig.~4 of \citet{2011MNRAS.411..162G}. Solid
  disks (\raisebox{-.2ex}{\large\textbullet}) mark the depth of the excess
  wing signature in \ion{He}{i}\,4471 (average of red and blue side, see also
  Fig.~\ref{fig:res_temp}) in the mean difference spectra. }
\end{figure*}
%
Achernar is also known to be a non-radial pulsator, with a pulsation frequency
of $F_1=0.775$\,c/d, which is both seen in spectroscopy and photometry
\citep[][and references therein]{2003A&A...411..229R}. A somewhat more
detailed picture was obtained by \citet{2011MNRAS.411..162G} based on
satellite photometry. They found the above mentioned period to be stable in
value and phase, but in amplitude being correlated with the circumstellar
activity state, i.e., when circumstellar emission is present, the amplitude is
high.  A stronger, secondary frequency $F_2$, at a value about 10\% lower, was
found to be present only at times of circumstellar activity, which is not
atypical of Be stars \citep{1998ASPC..135..348S}.

{\begin{table}[b]
\caption{\label{tab:obs}Observations used in this work}
\begin{center}\small
\begin{tabular}{llllr} 
Date range       & Instrument & Coverage &Resolv.& \# of\\
                 &            & [nm]    & power & spectra \\
\hline\\[-2ex]
Jan.\ 1999       & FEROS      & 370--890  & 48\,000 & 20 \\
Jun./Jul.\ 1999  & HEROS      & 350--860  & 20\,000 & 43 \\
Jan.\ 2000       & FEROS      & 370--890  & 48\,000 & 44 \\
Oct./Nov.\ 2001  & FEROS      & 370--890  & 48\,000 & 16 \\
Sep.\ 2002       & FEROS      & 370--890  & 48\,000 &  3 \\
Sep.--Dec.\ 2003 & HARPS      & 380--690  & 115\,000 & 11 \\
Jun.\ 2004       &  UVES      & 320--950  & 60\,000 & 17 \\
Sep./Oct.\ 2004  & HARPS      & 380--690  & 115\,000 & 6  \\
Jan.\ 2006       &  UVES      & 380--500  & 60\,000 & 60 \\
Jun.--Dec.\ 2006 & FEROS      & 370--890  & 48\,000 & 45 \\
\end{tabular}
\end{center}

\end{table}}

\section{Observations}

High-quality echelle spectra are required for the intended analysis. The data
sets available for this study (see Table~\ref{tab:obs}) were obtained with the
instruments
HEROS \citep{1995JAD.....1....3S}, FEROS \citep{1999Msngr..95....8K}, UVES
\citep{2000SPIE.4008..534D}, and HARPS \citep{2003Msngr.114...20M}.
Spectra taken after the year 2000 are available from the ESO Science Archive
Facility. HEROS technical parameters, observing procedures, and data reduction
are described by \citet{1995JAD.....1....3S}, UVES and FEROS data were reduced
with the standard data reduction systems provided by ESO, for HARPS the
reduced data were obtained from the archive.  Some spectra were discarded for
quality reasons or unsuitable observing modes, e.g., HARPS spectra taken with
iodine cell. {\bf See Appendix \ref{app:A} for a discussion of stability over
several seasons and instruments.}
H$\alpha$ equivalent widths were then measured in each spectrum
(Fig.~\ref{fig:EW}).  For the analysis of high quality differences between the
seasons, the spectra in each run listed in Table~\ref{tab:obs} were
averaged. {\bf Unity was added to the difference spectra, so that the
  continuum has a value of one instead of zero.}

\begin{figure}[t]
\begin{center}
\includegraphics[angle=0,width=8cm,clip]{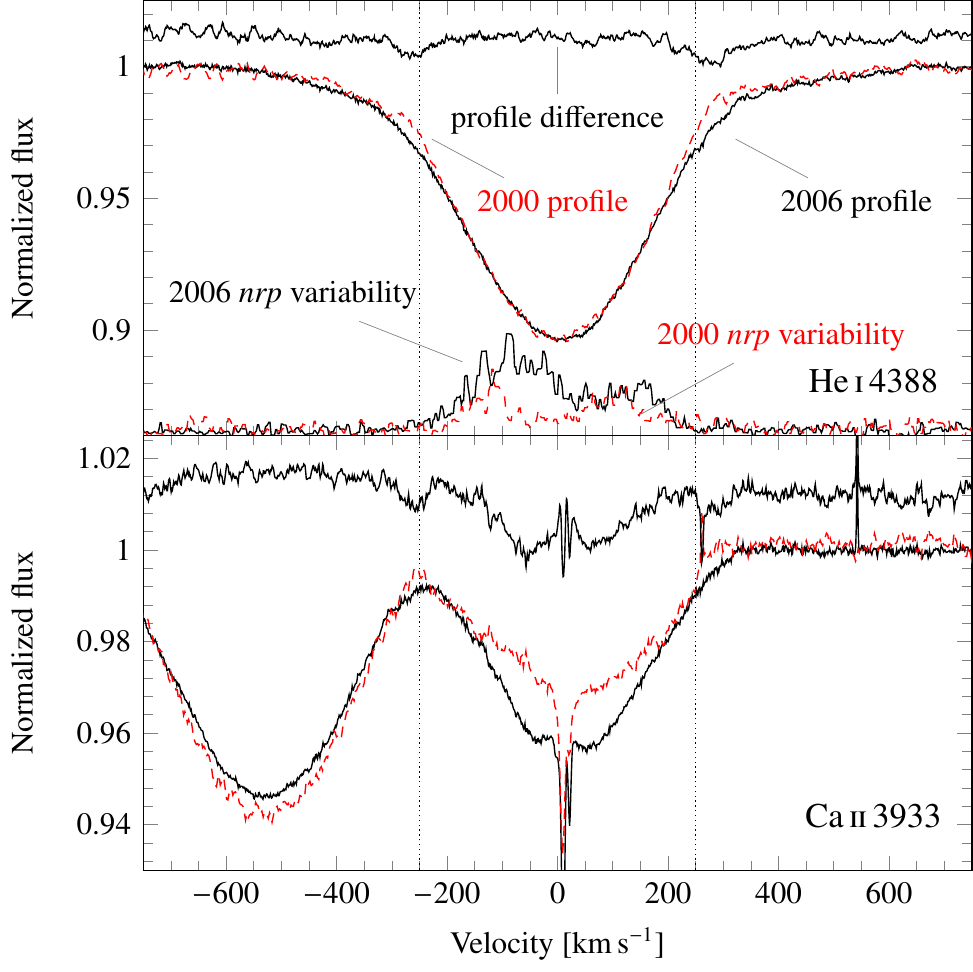}%
\end{center}
\caption[xx]{\label{fig:cqe} FEROS 2000 (dashed, red in online version) and
  UVES 2006 (solid) spectra and their differences. Shown are a purely
  photospheric line (\ion{He}{i}\,4388) and a line with circumstellar
  contribution (\ion{Ca}{ii}\,3933). The temporal variance spectra for the
  2000 and 2006 seasons are shown for \ion{He}{i}\,4388 as well (arbitrary
  scaling).  In neither season the pulsational power reached beyond
  $\pm250$\,km\,s$^{-1}$ (dotted lines). The CQE in \ion{Ca}{ii}\,3933 is
  embedded in a shell absorption (plus the interstellar line), but the
  additional absorption at high velocity is photospheric in both lines.  }
\end{figure}

\begin{figure}[t]
\begin{center}
\includegraphics[angle=0,width=8.cm,clip]{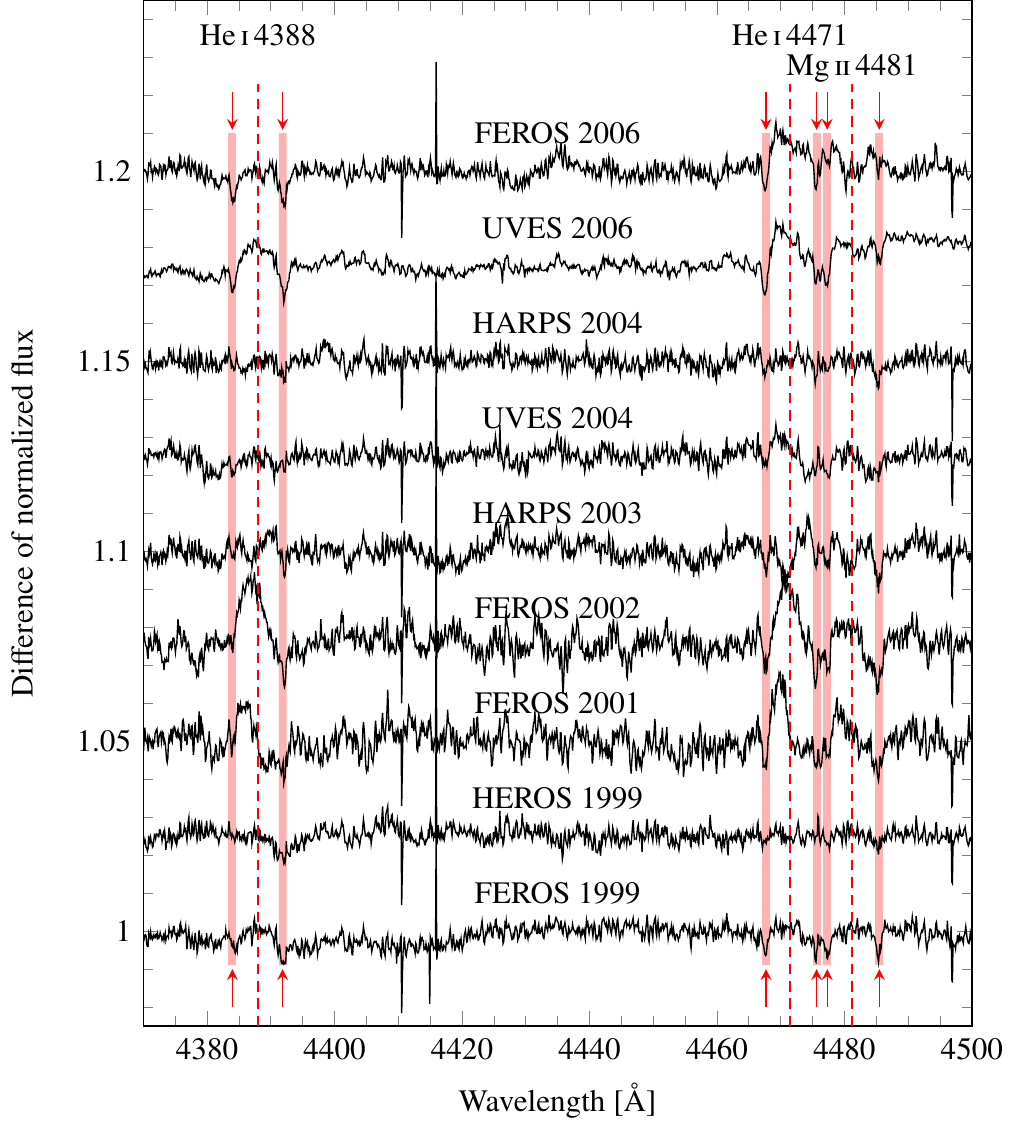}%
\end{center}
\caption[xx]{\label{fig:res_temp} Temporal evolution of the differences
  w.r.t.\ the pure photosphere state of 2000 (see as well
  Fig.~\ref{fig:EW}). Data were averaged for each observing epoch. In seasons
  with only few observing epochs (FEROS 2001, 2002, and HARPS 2003) residual
  pulsational profile distortions are clearly seen in the line cores.
  \ion{Mg}{ii}\,4481 shows circumstellar absorption in HARPS 2003, UVES 2006,
  and FEROS 2006, similar to the CQE seen in Fig.~\ref{fig:cqe}. Rest
  wavelengths of spectral lines are indicated with a dashed line, excess wing
  signature depression marked with arrows.}
\end{figure}

\section{Variations Relative to the Diskless State}

\subsection{Circumstellar Contributions}

Strong emission lines have never been observed in Achernar \citep[H$\alpha$
  peak height always $\lesssim1.5\,F_{\rm cont}$, see][]{2006A&A...446..643V},
but even these completely vanished between July and October 1999: Spectra
taken with {\sc FEROS} in January, 1999 and May--July, 1999 with HEROS show
weak emission in H$\alpha$ only, which had fully disappeared when the star was
observed in January, 2000 with FEROS again. The independent FEROS observations
by \citet{2006A&A...446..643V} were taken in October 1999 and confirm the
absence of any detectable disk. Taking the spectra of Jan.\ 2000 as
photospheric reference, one can construct the differences to test
circumstellar contribution or photospheric change against this reference.  By
the end of 2001 a weak disk had started forming (although in EW,
Fig.~\ref{fig:EW}, the new emission is balanced by simultaneous shell
absorption, it is clearly seen in the difference spectrum{\bf, see appendix
  \ref{app:B}}), that peaked in 2002, was almost gone again by the end of
2003, and had completely vanished by mid 2004. Thereafter there are no spectra
until 2006, by when a fully developed disk was present.  It is interesting to
compare this evolution to the report by \citet{2011MNRAS.411..162G}: The
photometric amplitudes of the two observed frequencies follow the same pattern
(see Fig.~\ref{fig:EW}).

The spectra taken in 2006, towards the end of a disk formation phase,
i.e., when the disk had reached maximum emission, show some properties normally
only seen in shell stars. 
The inclination values given by modeling ($\sim78^\circ$ by
\citealt{2012A&A...545A.130D}, $\sim65^\circ$ by
\citealt{2007ApJ...671L..49C}) are rather large, but not equatorial,
suggesting that Achernar might be a transition case between clear shell and
clear non-shell star.
Indicators for a shell-star nature of Achernar are 1) central quasi-emission
components \citep[CQE,][]{1999A&A...348..831R}, easily seen in
\ion{Ca}{ii}\,3933 (see Fig.~\ref{fig:cqe}) and \ion{Mg}{ii}\,4481, and 2)
differences from the photospheric spectrum of 1999/2000 below zero in the
center of lines most easily affected by circumstellar material.  The
circumstellar nature of these signatures is corroborated by comparison with
the modeled differences below, which, not counting the Balmer emission, do not
reproduce well the core of these, and only these, lines.

\subsection{Photospheric Variations}

The largest circumstellar variations are present in 2006. Apart from clear
double peaked emission and shell absorption contributions in the usual lines
(Balmer, lowly ionized metals), however, there is a third type of difference,
which is not restricted to lines typically formed in the circumstellar
environment, but is seen in purely photospheric lines, too, like the
\ion{He}{i} lines in the blue part of the spectrum (see Fig.~\ref{fig:cqe} for
\ion{He}{i}\,4388): In the UVES 2006 data, that line shows a shallower core,
but excess wings, not just deeper, but reaching out a bit farther than in
2000. In other words, the photospheric line is broader in 2006 than it is in
2000. The excess wing signature is as well seen in \ion{Ca}{ii}\,3933 in
Fig.~\ref{fig:cqe}, at higher velocity than the circumstellar shell
absorption.

\begin{figure*}[t]
\begin{center}
\includegraphics[angle=0,width=17cm,clip]{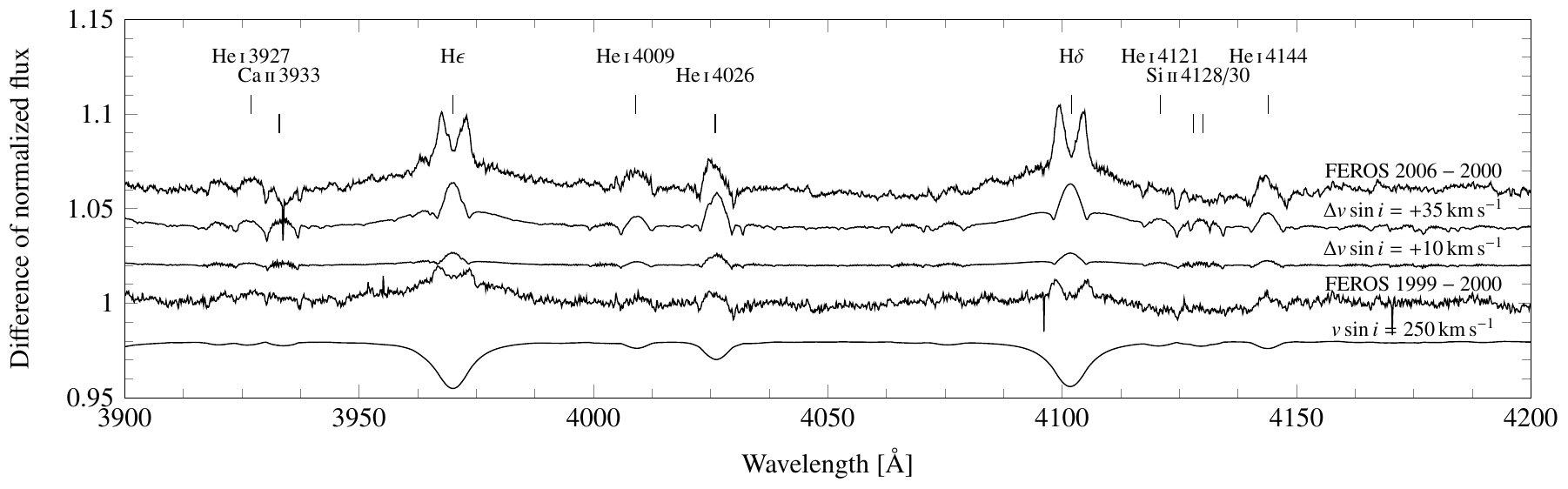}%
\end{center}
\caption[xx]{\label{fig:concept}As a proof of concept for a variable $v \sin
  i$ parameter, the differences of the FEROS 1999 and the FEROS 2006 to the
  FEROS 2000 spectrum are shown. Below is the modeled spectrum (scaled by a
  factor of 1/15), calculated with a fixed $T_{\rm eff}=15\,000$\,K,
  $\beta=0.25, v \sin i=250$\,km\,s$^{-1}$. The model differences of $\Delta v
  \sin i=+35$\,km\,s$^{-1}$ and $\Delta v \sin i=+10$\,km\,s$^{-1}$ are shown
  between the observed ones.
}
\end{figure*}

The pattern visible in Fig.~\ref{fig:cqe} is not only seen in all \ion{He}{i}
and the \ion{Ca}{ii} lines, but also in \ion{Mg}{ii}\,4481 (see
Fig.~\ref{fig:res_temp}), \ion{Si}{ii}\,4128/32 doublet (see
Fig.~\ref{fig:concept}), and as well in \ion{C}{ii}\,4267, \ion{Fe}{ii}\,5169,
\ion{Si}{ii}\,6347/71, or, in short, in basically all lines detected in the
spectrum with a depth larger than a few percent, with the notable exception of
\ion{Si}{iii}\,4553. The changes of the Balmer wings seen in
Fig.~\ref{fig:concept} are probably real as well; they are consistently
present in all Balmer lines except H$\alpha$.

This excess broadening of all photospheric lines changes on a time scale of
months to years. The temporal evolution is shown in Fig.~\ref{fig:res_temp}:
In all epochs the excess outer wings (see Fig.~\ref{fig:cqe}) are deeper than
in the diskless state observed with FEROS in 2000. The lines are narrowest
when the star has no circumstellar disk. Indeed, the differences are quite
small, there is almost no excess broadening in the HEROS 1999 data, and
quite little in the 2004 data, all with little to no sign of a disk.
The excess broadening is strongest, i.e., clearly seen deeper and wider line
wings, indicated in the difference spectra by ``dents'' at about $v
\pm275$\,km\,s$^{-1}$ of each line, when the disk strength is increasing or
steady (2001/2, 2006). When the disk strength is decreasing, the excess
broadening weakens (1999 and 2003). However, as 2004 data show, the excess
broadening can still be weakly present when no disk emission or shell absorption
can be detected anymore.

\subsection{Proof-of-Concept Modeling}

Since the variation is present in almost all photospheric lines in the same
manner (\ion{Si}{iii}\,4553 can be explained by its polar formation locus), a
change of the surface velocity field from epoch to epoch is a strong
possibility.  The variable pulsation amplitude, known from photometry, is one
possibility. However, an inspection of the individual spectra does not support
this hypothesis. The pulsational variations shown in an RMS variation analysis
are well contained inside $\pm250$\,km\,s$^{-1}$ for the 2006 season, and even
within $\pm220$\,km\,s$^{-1}$ for the 2000 season. Since that means there is
no variability outside that velocity, this cannot cause excess wings observed
between 250 and $300$\,km\,s$^{-1}$ (see Fig.~\ref{fig:cqe}).  Pairwise
differences between individual high-quality spectra do as well not show any
indication of the excess broadening to vary as part of the clearly seen
pulsational phase differences.

This variation of the excess broadening might instead come from the
photospheric rotational velocity field itself being different. As a proof of
concept for this hypothesis a set of photospheric model spectra was calculated
with Bruce3 \citep[see][for a description and
  references]{2013MNRAS.429..177R}. Model parameters were taken from published
sources: $R_{\rm pole} = 8\,{\rm R}_\odot$, $i=78^\circ$, $T_{\rm
  eff}=15\,000$\,K from \citet{2012A&A...545A.130D}, while the mass was
somewhat increased from their value of $M=6.1\,{\rm M}_\odot$ to $M=8.1\,{\rm
  M}_\odot$ to avoid the problem of interpolating equatorial local atmosphere
parameters outside the grid of model spectra available for Bruce3.  It should
be noted that the models used are strictly models for the photospheric
appearance only, so that a higher mass does not affect luminosity, etc., as it
would in a model including the stellar interior. Only the local gravity, the
geometric distortion, and the amount of gravity darkening are changed.
Standard gravity darkening with $\beta=0.25$ was adopted {\bf (see Appendix
  \ref{app:C} for a discussion of lower $\beta$)}. Spectra were computed
incrementing $v\sin i$ in steps of 5\,km\,s$^{-1}$, from 250 to
300\,km\,s$^{-1}$.

Figure~\ref{fig:concept} shows the model spectrum for $v\sin
i=250$\,km\,s$^{-1}$. At the scale of the figure, models computed with higher
$v\sin i$ look virtually identical. The difference spectra, however, exhibit
pronounced features, which are surprisingly close to the observed ones, given
that no sort of fitting beyond selecting an appropriate $\Delta v\sin i$ was
attempted. The strongest discrepancies between modeled and observed difference
spectra are the above identified circumstellar absorption/emission effects,
which are not included in the model.
Equatorially formed lines, such as singly ionized metals, are affected
differently than helium lines formed all over the star: In the equatorially
formed lines, the excess wings are very strong, but the core remains largely
unaltered, while both excess wings and line core change for \ion{He}{i}
lines. The more polar \ion{Si}{iii}\,4553 line does not change at a detectable
level at all. The Balmer line wings, finally, are less deep in the rapidly
rotating model due to the lower equatorial gravity. All this can be understood
in terms of the formation regions of the respective lines, as illustrated in
Fig.~6 of \citet{review}: The effects of varying $v \sin i$ are the strongest
at equatorial latitudes.  This as well explains why, contrary to classical
assumption, changes in $ v\sin i$ do affect the equivalent width.

The observed time-dependent line broadening
can be well explained by variations of the (equatorial) rotational velocity.

\section{Discussion and Conclusions}
High-quality spectroscopy of Achernar has shown a variable width of the
photospheric lines. The width correlates with the disk emission.
The variation can be interpreted as a change of the
stellar rotation speed, and a proof-of-concept model with $\Delta v \sin i
\leq 35$\,km\,s$^{-1}$ matches the observations already reasonably well
without parameter fitting. Such a velocity differential, on top of 95\% of
critical rotation, could make quite a difference for disk formation.
However, the absence of a detectable spin-up {\em before} the onset of disk
formation, marked by photometric amplitude increase in Fig.~\ref{fig:EW}, puts
constraints on models that first transport angular momentum upwards into the
outer photosphere, and then ``offload'' this angular momentum excess into a
circumstellar disk.  Instead, the excess width correlates with the {\em
  presence and decay} of the disk.

In a viscous disk scenario, ``decay of the disk'' actually means that disk
material is being reaccreted onto the central star. Even in build-up and
steady-state feeding phases, the very nature of viscous angular momentum
transport requires that some fraction of the material in the inner disk is
reaccreted.  A temporary accretional spin-up of the equatorial surface region
only is a possible hypothesis: The enhanced velocity is dissipated back into
deeper stellar regions after the reaccretion ceases, and the angular momentum
required for disk formation may be taken from such deeper seated regions
instead of the photospheric surface.

In any case, current models for stellar photospheres do not handle such a
situation well. For instance, instead of keeping $T_{\rm eff}$ constant with
the spin-up, one could as well keep either the polar temperature or the
luminosity constant; or implement an equatorially enhanced differential
rotation and correspondingly modified local gravity/temperature values instead
of a solid body spin-up; or change the values of gravity darkening and/or
equatorial temperature. However, to explore and lift these possible
degeneracies is well beyond the scope of this discovery report.

Since about January 2013 Achernar has been building up a new disk from almost
zero emission (T.\ Napole\~ao \& R. Marcon, priv.\ comm., 2013). With Achernar
being the closest and apparently brightest Be star, this offers a timely
opportunity to obtain dedicated observations, and to scrutinize whether the
spin-up is cause or consequence of disk formation. Similar findings for
other Be stars might be waiting for discovery in archival data.

\begin{acknowledgements}
RHDT acknowledges support from NASA award NNX12AC72G. ACC acknowledges support
from CNPq (307076/2012-1) and Fapesp (2010/19029-0). The referee has greatly
helped to improve the robustness against potential objections.
\end{acknowledgements}

\bibliographystyle{bibtex/aa} \bibliography{bibtex/alp}

\newcommand{\oneletter}[1]{#1}
\begin{thebibliography}{18}
\expandafter\ifx\csname natexlab\endcsname\relax\def\natexlab#1{#1}\fi

\bibitem[{{Ando}(1986)}]{1986A&A...163...97A}
{Ando}, H. 1986, \aap, 163, 97

\bibitem[{{Carciofi} {et~al.}(2007){Carciofi}, {Magalh{\~a}es}, {Leister},
  {Bjorkman}, \& {Levenhagen}}]{2007ApJ...671L..49C}
{Carciofi}, A.~C., {Magalh{\~a}es}, A.~M., {Leister}, N.~V., {Bjorkman}, J.~E.,
  \& {Levenhagen}, R.~S. 2007, \apjl, 671, L49

\bibitem[{{Dekker} {et~al.}(2000){Dekker}, {D'Odorico}, {Kaufer}, {Delabre}, \&
  {Kotzlowski}}]{2000SPIE.4008..534D}
{Dekker}, H., {D'Odorico}, S., {Kaufer}, A., {Delabre}, B., \& {Kotzlowski}, H.
  2000, in SPIE Conf.\ Ser., ed. M.~{Iye} \& A.~F. {Moorwood}, Vol. 4008, 534

\bibitem[{{Domiciano de Souza} {et~al.}(2012){Domiciano de Souza}, {Hadjara},
  {Vakili}, {Bendjoya}, {Millour}, {Abe}, {Carciofi}, {Faes}, {Kervella},
  {Lagarde}, {Marconi}, {Monin}, {Niccolini}, {Petrov}, \&
  {Weigelt}}]{2012A&A...545A.130D}
{Domiciano de Souza}, A., {Hadjara}, M., {Vakili}, F., {et~al.} 2012, \aap,
  545, A130

\bibitem[{{Domiciano de Souza} {et~al.}(2003){Domiciano de Souza}, {Kervella},
  {Jankov}, {Abe}, {Vakili}, {di Folco}, \& {Paresce}}]{2003A&A...407L..47D}
{Domiciano de Souza}, A., {Kervella}, P., {Jankov}, S., {et~al.} 2003, \aap,
  407, L47

\bibitem[{{Goss} {et~al.}(2011){Goss}, {Karoff}, {Chaplin}, {Elsworth}, \&
  {Stevens}}]{2011MNRAS.411..162G}
{Goss}, K.~J.~F., {Karoff}, C., {Chaplin}, W.~J., {Elsworth}, Y., \& {Stevens},
  I.~R. 2011, \mnras, 411, 162

\bibitem[{{Haubois} {et~al.}(2012){Haubois}, {Carciofi}, {Rivinius}, {Okazaki},
  \& {Bjorkman}}]{2012ApJ...756..156H}
{Haubois}, X., {Carciofi}, A.~C., {Rivinius}, T., {Okazaki}, A.~T., \&
  {Bjorkman}, J.~E. 2012, \apj, 756, 156

\bibitem[{{Kaufer} {et~al.}(1999){Kaufer}, {Stahl}, {Tubbesing},
  {N{\o}rregaard}, {Avila}, {Francois}, {Pasquini}, \&
  {Pizzella}}]{1999Msngr..95....8K}
{Kaufer}, A., {Stahl}, O., {Tubbesing}, S., {et~al.} 1999, The Messenger, 95, 8

\bibitem[{{Mayor} {et~al.}(2003){Mayor}, {Pepe}, {Queloz}, {Bouchy},
  {Rupprecht}, {Lo Curto}, {Avila}, {Benz}, {Bertaux}, {Bonfils}, {Dall},
  {Dekker}, {Delabre}, {Eckert}, {Fleury}, {Gilliotte}, {Gojak}, {Guzman},
  {Kohler}, {Lizon}, {Longinotti}, {Lovis}, {Megevand}, {Pasquini}, {Reyes},
  {Sivan}, {Sosnowska}, {Soto}, {Udry}, {van Kesteren}, {Weber}, \&
  {Weilenmann}}]{2003Msngr.114...20M}
{Mayor}, M., {Pepe}, F., {Queloz}, D., {et~al.} 2003, The Messenger, 114, 20

\bibitem[{{Neiner} {et~al.}(2013){Neiner}, Mathis, Saio, \& Lee}]{Coralie_2013}
{Neiner}, C., Mathis, S., Saio, H., \& Lee, U. 2013, in Progress in Physics of
  the Sun and Stars: A New Era in Helio- and Asteroseismology, ed.
  H.~{Shibahashi} \& A.~E. {Lynas-Gray}, ASPC, in press

\bibitem[{{Rivinius} {et~al.}(2003){Rivinius}, {Baade}, \& {{\v
  S}tefl}}]{2003A&A...411..229R}
{Rivinius}, T., {Baade}, D., \& {{\v S}tefl}, S. 2003, \aap, 411, 229

\bibitem[{{Rivinius} {et~al.}(2013{\natexlab{a}}){Rivinius}, Carciofi, \&
  Martayan}]{review}
{Rivinius}, T., Carciofi, A.~C., \& Martayan, C. 2013{\natexlab{a}}, \aapr,
  submitted

\bibitem[{{Rivinius} {et~al.}(2013{\natexlab{b}}){Rivinius}, {Townsend},
  {Kochukhov}, {{\v S}tefl}, {Baade}, {Barrera}, \&
  {Szeifert}}]{2013MNRAS.429..177R}
{Rivinius}, T., {Townsend}, R.~H.~D., {Kochukhov}, O., {et~al.}
  2013{\natexlab{b}}, \mnras, 429, 177

\bibitem[{{Rivinius} {et~al.}(1999){Rivinius}, {{\v S}tefl}, \&
  {Baade}}]{1999A&A...348..831R}
{Rivinius}, T., {{\v S}tefl}, S., \& {Baade}, D. 1999, \aap, 348, 831

\bibitem[{{Rogers} {et~al.}(2013){Rogers}, {Lin}, {McElwaine}, \&
  {Lau}}]{2013ApJ...772...21R}
{Rogers}, T.~M., {Lin}, D.~N.~C., {McElwaine}, J.~N., \& {Lau}, H.~H.~B. 2013,
  \apj, 772, 21

\bibitem[{{Stahl} {et~al.}(1995){Stahl}, {Kaufer}, {Wolf}, {Gang},
  {Gummersbach}, {Kovacs}, {Mandel}, {Rivinius}, {Szeifert}, \&
  {Zhao}}]{1995JAD.....1....3S}
{Stahl}, O., {Kaufer}, A., {Wolf}, B., {et~al.} 1995, Journal of Astronomical
  Data, 1, 3

\bibitem[{{{\v S}tefl} {et~al.}(1998){{\v S}tefl}, {Baade}, {Rivinius},
  {Stahl}, {Wolf}, \& {Kaufer}}]{1998ASPC..135..348S}
{{\v S}tefl}, S., {Baade}, D., {Rivinius}, T., {et~al.} 1998, in ASPC, Vol.
  135, A Half Century of Stellar Pulsation Interpretation, ed. P.~A. {Bradley}
  \& J.~A. {Guzik}, 348

\bibitem[{{Vinicius} {et~al.}(2006){Vinicius}, {Zorec}, {Leister}, \&
  {Levenhagen}}]{2006A&A...446..643V}
{Vinicius}, M.~M.~F., {Zorec}, J., {Leister}, N.~V., \& {Levenhagen}, R.~S.
  2006, \aap, 446, 643

\end{thebibliography}

\appendix
\section{Cross-Season and -Instrument Data}\label{app:A}
\setcounter{section}{2}\setcounter{figure}{0}
\begin{figure*}[t]
\begin{center}
\includegraphics[angle=0,width=17cm,clip]{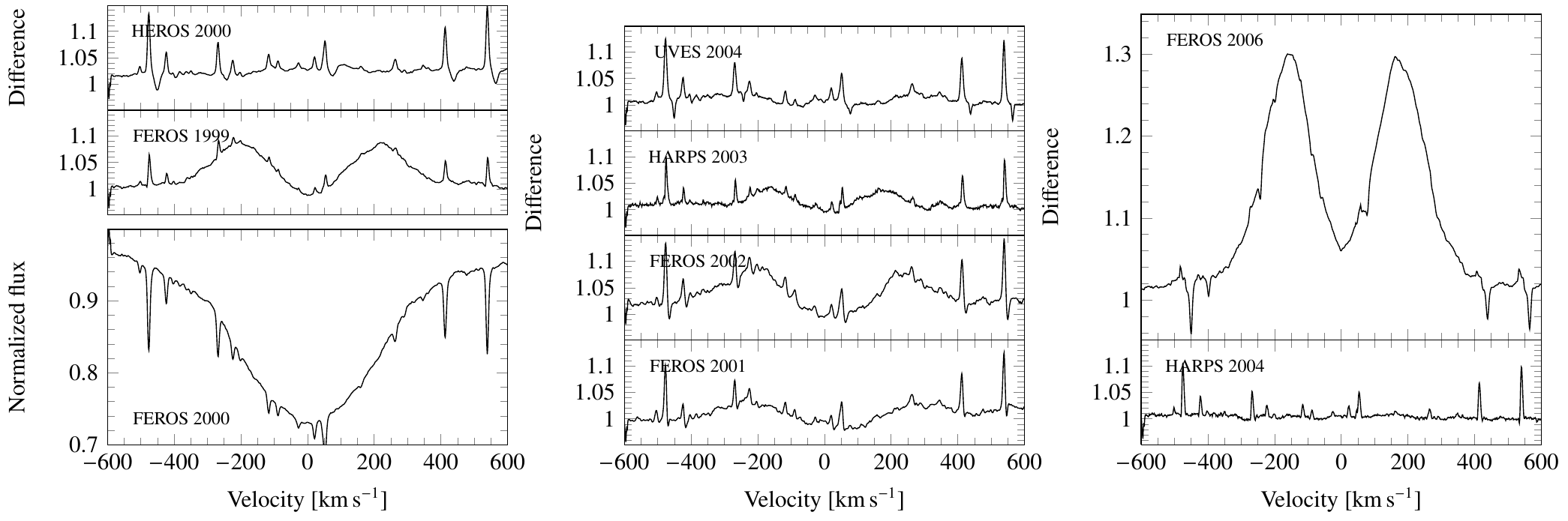}%
\end{center}
\caption[xx]{\label{fig:ha_appendix}H$\alpha$ difference profiles
  w.r.t.\ FEROS 2000 for all observing seasons. Telluric lines were not
  corrected for.
}
\end{figure*}
\setcounter{section}{1}

\setcounter{section}{3}\setcounter{figure}{0}
\begin{figure*}[t]
\begin{center}
\includegraphics[angle=0,width=17cm,clip]{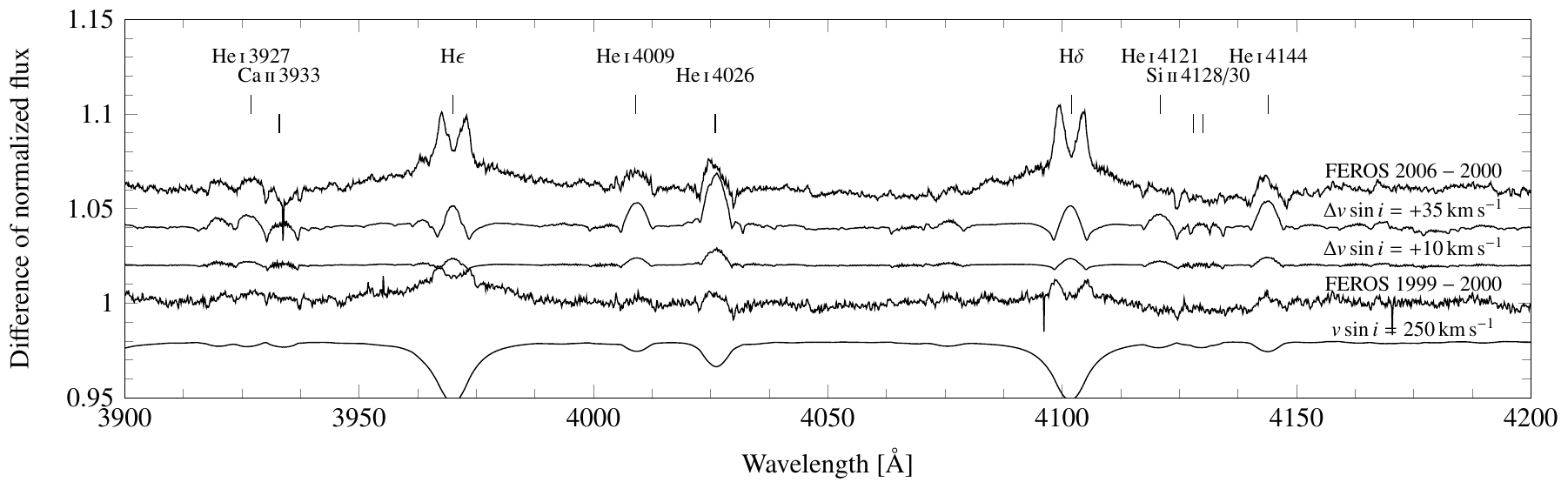}%
\end{center}
\caption[xx]{\label{fig:concept_appendix}Same as Fig.~\ref{fig:concept}, with
  models computed for a fixed $T_{\rm pole}=16900$\,K and $\beta=0.20$.
}
\end{figure*}
\setcounter{section}{1}

Using data from four different instruments, over about a decade, in order to
detect subtle effects might be problematic in terms of stability and
cross-instrumental effects. All are echelle spectrographs; HARPS, FEROS and
HEROS are attached by fiber-link, UVES is mounted on a gravity invariant
Nasmyth platform. The three ESO instruments are monitored, and partly
thermally controlled, for stability.  All spectra were used in the
heliocentric reference frame. The pixel sampling of the spectra is from
0.1\,\AA\ (HEROS) over 0.03\,\AA\ (FEROS) down to about 0.01\AA\ (UVES and
HARPS). This is much smaller than the width of the observed variations, which
are several tens of km\,s$^{-1}$ even for the most narrow ones, the high
velocity ``dents''. The observed line depths differ by almost 2\% of the
continuum, at a typical $S/N$ of the average spectra of about 1000.  The
observed effect is not systematically different between FEROS$-$FEROS and
OTHER$-$FEROS comparisons.

As the variations are well oversampled and have a consistent appearance and
evolution through all data sets, regardless of instrument, a
cross-instrumental/stability effect can be excluded.

\section{The \boldmath H$\alpha$\unboldmath\ Difference Spectra}
\label{app:B}
To allow an assessment of the circumstellar disk strength independent of the
H$\alpha$ equivalent width, the seasonal H$\alpha$ profiles are shown in
Fig.~\ref{fig:ha_appendix} in a similar format as given by
\citet{2006A&A...446..643V} in their Fig.~12.  In a shell star, the net effect
of a weakly developed disk on the overall H$\alpha$ equivalent width can
vanish if, as happened in Achernar in 2001, the contributions from line
emission and shell absorption just cancel one another. Note as well the
unusual Balmer-line CQE seen in the FEROS 2001 and somewhat more strongly in
UVES 2004 data.

\section{The Gravity Darkening Parameter}
\label{app:C}

In the proof-of-concept model a gravity darkening value of $\beta=0.25$ was
used, even though lower values of $\beta$ are typically derived
observationally. We note that $\beta=0.2$ was adopted by
\citet{2012A&A...545A.130D}, though not derived. The model we use can compute
spectra for different values of $\beta$, but the current formalism to compute
the corresponding stellar parameters, in particular $T_{\rm eff}$ and $L$, has
been derived with a fixed $\beta=0.25$. For this reason, modeling with a fixed
$T_{\rm eff}$ is only possible with $\beta=0.25$, a shortcoming that will be
overcome in a more detailed dedicated work. To demonstrate that the effect
does not vanish for a different $\beta$, we show models computed with constant
$T_{\rm pole}$ instead, for $\beta=0.20$
(Fig.~\ref{fig:concept_appendix}). Compared to the models shown in
Fig.~\ref{fig:concept}, the ``dents'' are unaltered in \ion{Ca}{ii}3933 and
the \ion{Si}{ii} lines, and slightly less pronounced in \ion{He}{i} lines, but
still clearly present in \ion{He}{i}\,3927,4009,4121,4144. The most
significant difference is that the Balmer line wings are not reproduced any
more (and Balmer lines are generally stronger), which, however, is an effect
dominated by keeping $T_{\rm pole}$ fixed rather than $T_{\rm eff}$, and
unrelated to the value of $\beta$.

\end{document}